\documentclass[12pt]{article}
\usepackage{graphicx}
\usepackage[utf8]{inputenc}
\usepackage[english
]{babel}
\def\baselinestretch{1.5}
\begin{document}
\begin{center}
\bf{Triangle geometry for qutrit states in the probability  representation.}\\
\end{center}
\bigskip

\begin{center} {\bf V. N. Chernega$^1$, O. V. Man'ko$^{1,2}$, V. I. Man'ko$^{1,3,4}$}
\end{center}

\medskip

\begin{center}
$^1$ - {\it Lebedev Physical Institute, Russian Academy of Sciences\\
Leninskii Prospect 53, Moscow 119991, Russia}\\
$^2$ - {\it Bauman Moscow State Technical University\\
The 2nd Baumanskaya Str. 5, Moscow 105005, Russia}\\
$^3$ - {\it Moscow Institute of Physics and Technology (State University)\\
Institutskii per. 9, Dolgoprudnyi, Moscow Region 141700, Russia}\\
$^4$ {\it Tomsk State University, Department of Physics\\
Lenin Avenue 36, Tomsk 634050, Russia}\\
Corresponding author e-mail: manko@sci.lebedev.ru
\end{center}

\section*{Abstract}
We express the matrix elements of the density matrix of the qutrit
state in terms of probabilities associated with artificial qubit
states. We show that the quantum statistics of qubit states and
observables is formally equivalent to the statistics of classical
systems with three random vector variables and three classical
probability distributions obeying special constrains found in this
study. The Bloch spheres geometry of qubit states is mapped onto
triangle geometry of qubits. We investigate the triada of Malevich's
squares describing the qubit states in quantum suprematism picture
and the inequalities for the areas of the squares for qutrit (spin-1
system). We expressed quantum channels for qutrit states in terms of
a linear transform of the probabilities determining the qutrit-state
density matrix.

\section{Introduction}  
The pure states of qutrit are described by a vector in the
three-dimensional Hilbert space~\cite{Diracbook}. The mixed states
of qutrit are described by the three-dimensional density
matrix~\cite{Landau}. The qutrit states can be realized as the
states of a spin-1 particle or as the states of the three-level
atom. The density matrix of the spin state in the spin tomographic
probability representation~\cite{DodPLA,OlgaJetp} is determined by a
fair probability distribution of spin projections on arbitrary
directions in the space called the spin tomogram. The von Neumann
entropy~\cite{VonNeumannbook32} of the qutrit state was
shown~\cite{ChernegaJRLR2013} to satisfy the entropic inequality,
which is the subadditivity condition analogous to the subadditivity
condition for bipartite systems of two qubits.
Recently~\cite{Malevich,70}, the triangle geometry of qubit states,
in which the density matrix of the spin-1/2 particle was associated
with the triada of Malevich's squares, was investigated. The areas
of the Malevich's squares are determined by three tomographic
probabilities of spin projections $m=1/2$ onto three perpendicular
directions in the space.

The aim of this work is to construct the triada of Malevich's
squares associated with the density matrix of qutrit states using
the approach connecting the qutrit states with the states of two
artificial qubits found in~\cite{ChernegaJRLR2013} and extra
artificial qubit associated with the permutation of the axes
$x\longleftrightarrow z$ in the three-dimensional space. We review
the probability description of qubit
states~\cite{Malevich,70,Marmoarx2016MarmoJPhysA2017} and derive
compact formulas for spin tomograms of these states. We use the
relation of quitrit states to the states of artificial qubits to
express the density matrix elements of the qutrit state in terms of
probabilities of the spin-1/2 projection.

This paper is organized as follows.

In Sec.~2, we review the quantum suprematism picture of spin-1/2
particle states suggested in~\cite{Malevich,70}. In Sec.~3, we
discuss the statistical properties of the quantum spin-1/2
observable. In Sec.~4, we consider the qutrit-state density matrix
and express its matrix elements in terms of probabilities of
spin-1/2 projections related to three artificial qubit states
connected with the given indivisible qutrit system. In Sec.~5, we
discuss the triangle geometry of the qutrit state and study the
inequalities for the tomographic probabilities determining the state
density matrix. We present our conclusions and prospectives in
Sec.~6.

\section{Qubits in the Quantum Suprematism Picture} 
The density matrix of qubit states is the Hermitian 2$\times$2
matrix $\rho$ satisfying the conditions $\rho^\dagger=\rho$,
$\mbox{Tr}\rho=1$, and $\rho\geq0$. This means that the density matrix has
two eigenvalues, which are nonnegative numbers $\lambda_1$ and
$\lambda_2$, with $\lambda_1+\lambda_2=1$. We consider the matrix
$~\rho=\left(\begin{array}{cc}
  \rho_{11}&\rho_{12}\\
  \rho_{21}&\rho_{22}\\
  \end{array}\right).$
The eigenvalues $\lambda_1$ and $\lambda_2$ of the density matrix
$\rho$ satisfy the equation
\begin{equation}\label{eq.2}
\left(\rho_{11}-\lambda\right)\left(\rho_{22}-\lambda\right)
  -\rho_{12}\rho_{21}=0.
\end{equation}
It was shown in~\cite{Marmoarx2016MarmoJPhysA2017} that the matrix
elements of the density matrix $\rho$ can be expressed within the
framework of the probability representation of qubit states in terms
of three probabilities $0\leq p_1,p_2,p_3\leq1$, namely,
\begin{equation}\label{eq.3}
\rho=\left(\begin{array}{cc}
p_3&p_1-ip_2-(1/2)+(i/2)\\
p_1+ip_2-(1/2)-(i/2)&1-p_3\\
\end{array}\right).
\end{equation}
In this expression, nonnegative probabilities $p_1$, $p_2$, and
$p_3$ are the probabilities of spin-1/2 projections $m=1/2$ onto
three perpendicular directions in the space, namely, $p_1$ is the
probability to have the spin projection along the $x$~direction,
$p_2$ is the probability to have the spin projection along the
$y$~direction, and $p_3$ is the probability to have the spin
projection along the $z$~direction. The eigenvalues of the density
matrix~(\ref{eq.3}) read
\begin{eqnarray}
\lambda_1=\frac{1}{2}+\left[\sum_{j=1}^3\left(p_j-\frac{1}{2}\right)^2\right]^{1/2},\qquad
\lambda_2=\frac{1}{2}-\left[\sum_{j=1}^3\left(p_j-\frac{1}{2}\right)^2\right]^{1/2}.
\label{eq.5}
\end{eqnarray}
The nonnegativity of the density matrix provides the
inequality~\cite{Marmoarx2016MarmoJPhysA2017} for three
probabilities $p_j$, namely,
\begin{equation}\label{eq.6}
\left(p_1-{1}/{2}\right)^2+\left(p_2-{1}/{2}\right)^2+\left(p_3-{1}/{2}\right)^2\leq {1}/{4}.
\end{equation}
This inequality is the nonnegativity condition for the density
matrix of the qubit state; it reflects the presence of quantum
correlations of the single-spin states.

There exists the geometrical interpretation of the introduced
parameters of the spin-state density matrix. The probabilities
$p_1$, $p_2$, and $p_3$ can be associated with a triangle on the
plane~\cite{Malevich,70}. The lengths $L_n$ $(n=1,2,3)$ of the
triangle sides are expressed in terms of the probabilities as
follows:
\begin{eqnarray}\label{eq.M2}
L_1=\left(2+2p_2^2-4p_2-2p_3+2p_3^2+2p_2
p_3\right)^{1/2},\nonumber\\
L_2=\left(2+2p_3^2-4p_3-2p_1+2p_1^2+2p_3
p_1\right)^{1/2},\\
L_3=\left(2+2p_1^2-4p_1-2p_2+2p_2^2+2p_1 p_2\right)^{1/2}.\nonumber
\end{eqnarray}
The probabilities $p_1$, $p_2$, and $p_3$ satisfy the inequality
\begin{equation}\label{eq.VB}
L_n+L_{n-1}>L_{n+1},\qquad n=1,2,3.
\end{equation}
Three squares with these sides and the areas $S_n=L_n^2$ were
introduced in~\cite{Malevich,70}; they were called the triada of
Malevich's squares. The area of triangle with the sides $L_n$ reads
\begin{equation}\label{eq.M7}
S_{\mbox{tr}}=({1}/{4})\big[\left(L_1+L_2+L_3\right)\left(L_1+L_2-L_3\right)
\left(L_2+L_3-L_1\right)\left(L_3+L_1-L_2\right)\big]^{1/2}.
\end{equation}
Usually, the density matrix (\ref{eq.3}) is associated with a point
in the Bloch ball. In the triangle geometry picture under
discussion, the density matrix is represented by the triada of
Malevich's squares. This means that we construct the invertible map
of any point in the Bloch ball onto the triangle with sides $L_n$
and the triada of Malevich's squares. The obvious inequalities for
the triangle sides give the inequalities for the probabilities
$p_1$, $p_2$, and $p_3$~(\ref{eq.VB}). These inequalities are
compatible with the condition~(\ref{eq.6}). The three squares
introduced in~\cite{Malevich,70} and called the triada of Malevich's
squares provide the quantum suprematism picture of the qubit
states.\footnote{We thank Dr. Tommaso Calarco for informing us about
available discussions of Malevich's square picture related to
quantum states of a single atom (private communication).} It is
worth noting that Zeilinger, emphasizing in~\cite{7'} the importance
in physics to make experiments as simple as possible and with the
smallest efforts, compared such approach with the creation of
Malevich's black square in the art.

The sum of areas of three Malevich's squares expressed in terms of
the probabilities $p_1$, $p_2$, and $p_3$ reads
\begin{equation}\label{eq.7}
S=2\left[3\left(1-p_1-p_2-p_3\right)+2p_1^2+2p_2^2+2p_3^2+p_1p_2+p_2p_3+p_3p_1\right].
\end{equation}
The sum satisfies the inequality
\begin{equation}\label{eq.8}
{3}/{2}\leq S<{9}/{2}.
\end{equation}
For classical system of three coins, an analogous suprematism
picture of Malevich's squares provides for this sum the domain
${3}/{2}\leq S\leq6$. The difference between numbers 9/2 and 6
reflects the difference of classical and quantum correlations in the
two systems -- qubit and three coins, though the states in both
cases are determined by three probabilities $p_1$, $p_2$, and $p_3$.

\section{Statistical Properties of Quantum Observable}
In this section, we discuss the properties of means of an observable
$A$ given by the Hermitian matrix
$~A_{j k}=\left(\begin{array}{cc}
A_{11}&A_{12}\\
A_{21}&A_{22}\end{array}\right).$
The mean values of the observable in the state with the density
matrix (\ref{eq.3}) read
\begin{equation}\label{3.2}
\langle A\rangle=\mbox{Tr} A\rho=p_3 A_{11}+(1-p_3)A_{22}+A_{12}
\big(p_1+i p_2-(1+i)/{2}\big)+A_{21}\big(p_1-i p_2
-({1-i})/{2}\big).
\end{equation}
This relation can be interpreted using the picture of three
classical random observables, which are described by three
probability distributions.

In fact, there are three probability vectors
\[
\vec{\cal P}_1=\left(\begin{array}{c}
p_1\\
1-p_1
\end{array}\right),\qquad \vec{\cal P}_2=\left(\begin{array}{c}
p_2\\
1-p_2
\end{array}\right),\qquad \vec{\cal P}_3=\left(\begin{array}{c}
p_3\\
1-p_3
\end{array}\right).
\]
For a spin-1/2 system, probabilities $(1-p_1)$, $(1-p_2)$, and
$(1-p_3)$ are the probabilities to have the spin-projection $m=-1/2$
along the axes $x$, $y$, and $z$, respectively. The matrix elements
of the matrix $A_{jk}$ $(j,k=1,2)$ can be considered as linear
functions of classical random variables, which take the real values
\begin{eqnarray}\label{3.4}
X_1=\frac{A_{12}+A_{21}}{2}, \quad
Y_1=\frac{i(A_{12}-A_{21})}{2},\quad
X_2=-\frac{A_{12}+A_{21}}{2},\quad Y_2=-\frac{i(A_{12}-A_{21})}{2}
\end{eqnarray}
and
\begin{equation}\label{3.5}
Z_1=A_{11}, \quad Z_2=A_{22}.
\end{equation}
The inverse relations are
\begin{equation}\label{3.5'}
A_{12}=X_1-iY_1,\qquad A_{11}=Z_1,\qquad A_{22}=Z_2,\quad
A_{21}=X_1+iY_1.
\end{equation}
Introducing the vector notation for the classical variables
\begin{equation}\label{3.6}
\vec X=\left(\begin{array}{c} X_1\\X_2\end{array}\right),\qquad
\vec Y=\left(\begin{array}{c} Y_1\\Y_2\end{array}\right),\qquad
\vec Z=\left(\begin{array}{c} Z_1\\Z_2\end{array}\right),
\end{equation}
we obtain the expression for the mean value of quantum observable
$\langle A\rangle$ in terms of the mean values of classical
observables $\vec X$, $\vec Y$, and $\vec Z$ of the form
\begin{equation}\label{3.7}
\langle A\rangle=\vec{\cal P}_1\vec X+\vec{\cal P}_2\vec Y+\vec{\cal P}_3\vec Z.
\end{equation}
Thus, the quantum relation for the mean value of the spin observable
$A$ in the state with the density matrix $\rho$ given by
(\ref{eq.3}) is presented as the sum of three classical means of
random variables $\vec X$, $\vec Y$, and $\vec Z$,
\begin{equation}\label{3.8}
\langle A\rangle =p_1X_1+(1-p_1)X_2+p_2Y_1+(1-p_2)Y_2+p_3Z_1+(1-p_3)Z_2.
\end{equation}
These observations provide a possibility to construct the model of
quantum observable $A$ using the classical observables $\vec X$,
$\vec Y$, and $\vec Z$.

In fact, for given arbitrary three real two-vectors $\vec X$, $\vec
Y$, and $\vec Z$ such that $X_1+X_2=Y_1+Y_2=0$, we construct the
Hermitian matrix $A_{j k}$ $(j,k=1,2)$ with matrix elements
(\ref{3.5'}). Since the density matrix~(\ref{eq.3}) is expressed in
terms of classical probability vectors $\vec {\cal P}_1$, $\vec
{\cal P}_2$, and $\vec {\cal P}_3$, the measurable quantum
observable $A$ has the mean value determined by classical
observables $\vec X$, $\vec Y$, $\vec Z$ and classical probability
distributions.

Quantumness of the model is formulated as inequality~(\ref{eq.6})
reflecting the condition for classical pro\-babilities $p_1$, $p_2$,
and $p_3$, and the definition of the second moment of quantum
observable $\langle A^2\rangle$ in terms of classical random
variables $\vec X$, $\vec Y$, and $\vec Z$,
\begin{eqnarray}
\mbox{Tr}\rho A^2&=&p_3Z_1^2-(1-p_3)Z_2^2+X_1^2+Y_1^2+2(Z_1+Z_2)\big[X_1(p_1-{1}/{2})+Y_1(p_2-{1}/{2})\big]\nonumber\\
&=&(Z_1+Z_2)\left[\vec X\vec{\cal P}_1+\vec Y\vec{\cal P}_2\right]+(X_1^2+Y_1^2)+ p_3(Z_1^2-Z_2^2)+Z_2^2.\label{3.9}
\end{eqnarray}
The constructed relations~(\ref{3.5'})--(\ref{3.8}) and
formulas~(\ref{3.8}) and (\ref{3.9}) for the quantum mean and
dispersion of any observable $A$, expressed in term of classical
random variables and classical probabilities, demonstrate that
quantum mechanics of qubits can be formulated using only standard
ingredients of classical probability theory. We conjecture that
quantum mechanics of any qudit system can also be formulated using
only classical random variables and classical probability
distributions. The difference from classical statistical mechanics
is expressed by specific inequalities for classical probability 
distributions, reflecting hidden correlations in quantum systems
analogous to (\ref{eq.6}) for qubits.

\section{Qutrit in the Probability Representation}
The tomographic probability distribution for the spin-1 system for a
minimum number of probabilities can be described by eight
parameters, which are spin projections $m=+1,0$ onto four
directions; these probabilities are discussed
in~\cite{Marmoarx2016MarmoJPhysA2017}. In this section, we develop
another approach to associate the density matrix of the qutrit state
with probabilities determining the states of artificial qubits.

We follow the approach applied to get a new entropic subadditivity
condition for the qutrit state suggested in~\cite{ChernegaJRLR2013}.
The density matrix of the spin-1 system is given by the matrix
$\rho$, such that $\rho^\dagger=\rho$, $\mbox{Tr}\rho=1$, and $\rho\geq0$;
it reads
\begin{equation}\label{eq.9}
\rho=\left(\begin{array}{ccc}
\rho_{11}&\rho_{12}&\rho_{13}\\
\rho_{21}&\rho_{22}&\rho_{23}\\
\rho_{31}&\rho_{32}&\rho_{33}\\
\end{array}\right).
\end{equation}
Applying the tool to consider the matrix $\rho$ as the 3$\times$3
block matrix in the 4$\times$4 density matrix of two qubits with
zero fourth column and zero fourth row, we obtain two qubit-state
density matrices of the artificial qubits using the partial tracing
procedure. The 2$\times$2 matrices are
\begin{equation}\label{eq.10}
\rho(1)=\left(\begin{array}{cc}
\rho_{11}+\rho_{22}&\rho_{13}\\
\rho_{31}&\rho_{33}\\
\end{array}\right),\qquad \rho(2)=\left(\begin{array}{cc}
\rho_{11}+\rho_{33}&\rho_{12}\\
\rho_{21}&\rho_{22}\\
\end{array}\right).
\end{equation}
For these two qubit-state density matrices, we have the expressions
in the probability representation in terms of probabilities
$p_{1,2,3}^{(k)}$, $k=1,2$, of the form
\begin{equation}\label{eq.11}
\rho(k)=\left(\begin{array}{cc}
p_3^{(k)}&p_1^{(k)}-ip_2^{(k)}-({1}/{2})+({i}/{2})\\
p_1^{(k)}+ip_2^{(k)}-({1}/{2})-({i}/{2})&1-p_3^{(k)}\\
\end{array}\right), \quad k=1,2.
\end{equation}
This means that a part of the matrix elements of the density matrix
$\rho$ is expressed in terms of the probabilities $p_j^{(k)},
~k=1,2,\,j=1,2,3$, namely,
\begin{equation}\label{eq.12}
\rho_{11}=p_3^{(2)}-\big(1-p_3^{(1)}\big),\qquad \rho_{22}=1-p_3^{(2)}, \qquad \rho_{33}=1-p_3^{(1)}.
\end{equation}
For off-diagonal matrix elements, we have
\begin{eqnarray}
\rho_{12}=p_1^{(2)}-ip_2^{(2)}-({1}/{2})+({i}/{2}),\qquad
\rho_{21}=\rho_{12}^{\ast},\label{eq.13}\\
\rho_{13}=p_1^{(1)}-ip_2^{(1)}-({1}/{2})+({i}/{2}),\qquad \rho_{31}=\rho_{13}^{\ast}.\label{eq.14}
\end{eqnarray}

To obtain an explicit expression for the matrix element $\rho_{23}$
in terms of probabilities, we consider the density matrix of the
state where we use the permutation of axes $x\leftrightarrow z$;
this means that we use another qutrit state. For a three-level atom,
we use the permutation of the ground state level and maximum excited
energy level; in such a case, we have the extra qubit with the
density matrix
\begin{equation}\label{eq.15}
\rho(3)=\left(\begin{array}{cc}
\rho_{33}+\rho_{11}&\rho_{32}\\
\rho_{23}&\rho_{22}\\
\end{array}\right).
\end{equation}
The probabilities $p_j^{(3)}$ for this artificial qubit state read
\begin{equation}\label{eq.16}
p_3^{(3)}=\rho_{11}+\rho_{33}=p_3^{(2)},\qquad p_1^{(3)}-i p_2^{(3)}
-(1-i)/2=\rho_{32},\qquad \rho_{23}=\rho_{32}^{\ast}.
\end{equation}
Thus, we provide the final expression of the qutrit density matrix
$\rho$ in terms of eight parameters -- probabilities $p_1^{(1)}$,
$p_2^{(1)}$, $p_3^{(1)}$, $p_1^{(2)}$,  $p_2^{(2)}$, $p_3^{(2)}$,
$p_1^{(3)}$, and $p_2^{(3)}$. The density matrix $\rho$ is
\begin{equation}\label{eq.17}
\rho=\left(\begin{array}{ccc}
p_3^{(2)}+p_3^{(1)}-1&~p_1^{(2)}-i p_2^{(2)}-(1-i)/2&~p_1^{(1)}+i p_2^{(1)}-(1+i)/2\\
p_1^{(2)}+i p_2^{(2)}-(1+i)/2&~1-p_3^{(2)}&~p_1^{(3)}+i p_2^{(3)}-(1+i)/2\\
p_1^{(1)}-i p_2^{(1)}-(1-i)/2&~p_1^{(3)}-i p_2^{(3)}-(1-i)/2&~1-p_3^{(1)}\\
\end{array}\right).
\end{equation}
The parameters $p_j^{(k)}$, $k,j=1,2,3$ must satisfy the inequalities
\begin{equation}\label{intq}
\sum_{j=1}^3(p_j^{(k)}-1/2)^2\leq 1/4.
\end{equation}
In addition to these inequalities, one has the cubic inequality
$\mbox{det}\,\rho\geq0$ and the quadratic inequality like
\begin{equation}\label{eq.19}
(1-p_3^{(2)})\,(1-p_3^{(1)})-|p_1^{(3)}+ip_2^{(3)}-({1+i})/{2}|^2\geq
0.
\end{equation}
To check all the inequalities, one needs to provide the
probabilities of spin-projections $m=+1/2$ onto three perpendicular
directions for the three artificial qubits. For the two qubits, the
directions are given by the axes $x$, $y$, and $z$, and for the
third qubit the direction corresponds to the permutation of the
first and the third directions, $x\leftrightarrow z$.

The density matrix $\rho$ can be rewritten in the form
\begin{equation}\label{A}
\rho=\left(\begin{array}{ccc}
p_3^{(2)}+p_3^{(1)}-1&p^{(2)\ast}-\gamma^\ast&p^{(1)}-\gamma\\
p^{(2)}-\gamma&1-p_3^{(2)}&p^{(3)}-\gamma\\
p^{(1)\ast}-\gamma^\ast&p^{(3)\ast}-\gamma^\ast&1-p_3^{(1)}\end{array}\right),
\end{equation}
where the complex numbers $p^{(k)}$ are
$~p^{(k)}=p_1^{(k)}+ip_2^{(k)}$, $k=1,2,3$, 
and $\gamma=(1+i)/2$. Then we express the purity of the qutrit state
$\mu=\mbox{Tr}\rho^2$ in terms of three classical probabilities
$p_j^{(k)}$, $j,k=1,2,3$,
\begin{equation}\label{C}
\mu=\big(p_3^{(2)}+p_3^{(1)}-1\big)^2+\big(1-p_3^{(2)}\big)^2+\big(1-p_3^{(1)}\big)^2+
2\big[|p^{(1)}-\gamma|^2+|p^{(2)}-\gamma|^2+|p^{(3)}-\gamma|^2\big].
\end{equation}
The nonnegativity condition of the density matrix
$\mbox{det}\,\rho\geq0$ yields the inequality for the probabilities
$p_j^{(k)}$, which looks like the inequality for the cubic
polynomial,
\begin{eqnarray}
&&\big(p_3^{(2)}+p_3^{(1)}-1\big)\big(1-p_3^{(2)}\big)\big(1-p_3^{(1)}\big)+
\big(p^{(2)}-\gamma\big)\big(p^{(1)}-\gamma\big)\big(p^{(3)\ast}-\gamma^\ast\big)\nonumber\\
&&+\big(p^{(2)\ast}-\gamma^\ast\big)\big(p^{(1)\ast}-\gamma^\ast\big)\big(p^{(3)}-\gamma\big)-
|p^{(1)\ast}-\gamma^\ast|^2\big(1-p_3^{(2)}\big)
-|p^{(2)}-\gamma|^2\big(1-p_3^{(1)}\big)\nonumber\\
&&-|p^{(3)}-\gamma|^2\big(p_3^{(2)}+p_3^{(1)}-1\big)\geq0.
\label{D}
\end{eqnarray}

The obtained inequalities (\ref{intq}), (\ref{eq.19}), and (\ref{D})
are quantum characteristics of the qutrit state expressed in terms
of classical probabilities $p_j^{(k)}$. One can extend the model of
qubit state based on the properties of classical random variables
$\vec X$, $\vec Y$, and $\vec Z$~(\ref{3.6}) to the case of the
qutrit state. We consider random classical variables $\vec X^{(k)}$,
$\vec Y^{(k)}$, and $\vec Z^{(k)}$, $k=1,2,3$ with the probability
distributions given by the vectors $\vec{\cal P}_1^{(k)}$,
$\vec{\cal P}_2^{(k)}$, and $\vec{\cal P}_3^{(k)}$.

If inequalities (\ref{intq}), (\ref{eq.19}), and (\ref{D}) are not
valid, the system properties correspond to the behavior of sets of
classical ``coins.'' Namely, quantum correlations are described by
inequalities (\ref{intq}), (\ref{eq.19}), and (\ref{D}). Thus, we
obtain new inequalities for qutrit states, which are entropic
inequalities for the probability vectors $\vec{\cal P}_j^{(k)}$,
$j,k=1,2,3$. For example, the inequality for relative entropy
\begin{equation}\label{E}
\sum_{j=1}^2p_j^{(k)}\ln \big(\,p_j^{(k)}/ p_j^{(k')}\big)\geq 0,\qquad k,k'=1,2,3,
\end{equation}
is valid for two arbitrary probability distributions. Since the
probabilities are expressed in terms of the density matrix elements
of the qutrit state, one has new entropic inequalities for the
qutrit-state density matrix; it is just inequality~(\ref{E}), which
provides the entropic inequality for the matrix elements of the
qutrit-state density matrix.

For example, one has the new relative-entropy inequality for the
matrix elements of the qutrit-state density matrix
\begin{equation}\label{G}
\frac{1}{2}\left(\rho_{12}+\rho_{21}+1\right)\ln
\left[\frac{\rho_{12}+\rho_{21}+1}{\rho_{13}+\rho_{31}+1}\right]
+\frac{1}{2}\left[i(\rho_{12}-\rho_{21})-1\right]
\ln\left[\frac{i(\rho_{12}-\rho_{21})-1}{i(\rho_{13}-\rho_{31})-1}\right]\geq0.
\end{equation}
An arbitrary permutation of indices $1,2,3$ in (\ref{G}) yields
another entropic inequality for the matrix elements of the
qutrit-state density matrix.

Now we discuss the geometric picture of the qutrit state using the
quantum suprematism approach. Each qubit state is visualized in
terms of the triada of Malevich's squares. The qutrit state, as we
have shown, is mapped onto three qubit states, which are described
by probabilities $p_j^{(k)}$, $j,k=1,2,3$. Among these nine
probabilities, eight are independent, but $p_3^{(3)}=p_3^{(2)}$.
Thus, the state can be described by three triadas of Malevich's
squares; see Fig.~1. Three sets of the Malevich's squares are
determined by the probabilities $p_1^{(1)}$, $p_2^{(1)}$,
$p_3^{(1)}$, $p_1^{(2)}$, $p_2^{(2)}$, $p_3^{(2)}$, $p_1^{(3)}$,
$p_2^{(3)}$, and $p_3^{(3)}$, where $p_3^{(2)}=p_3^{(3)}$. The sums
of the areas of the triadas of Malevich's squares are given by
(\ref{eq.7}). For each of the three triadas of Malevich's squares,
one has the inequality for the sums of the areas given by
(\ref{eq.8}). The inequality reflects the presence of quantum
correlations between the artificial qubits in the single-qutrit
state. During the time evolution of qutrit states, the inequalities
for the areas of Malevich's squares are respected.


\begin{figure}
 \includegraphics[width=50mm]{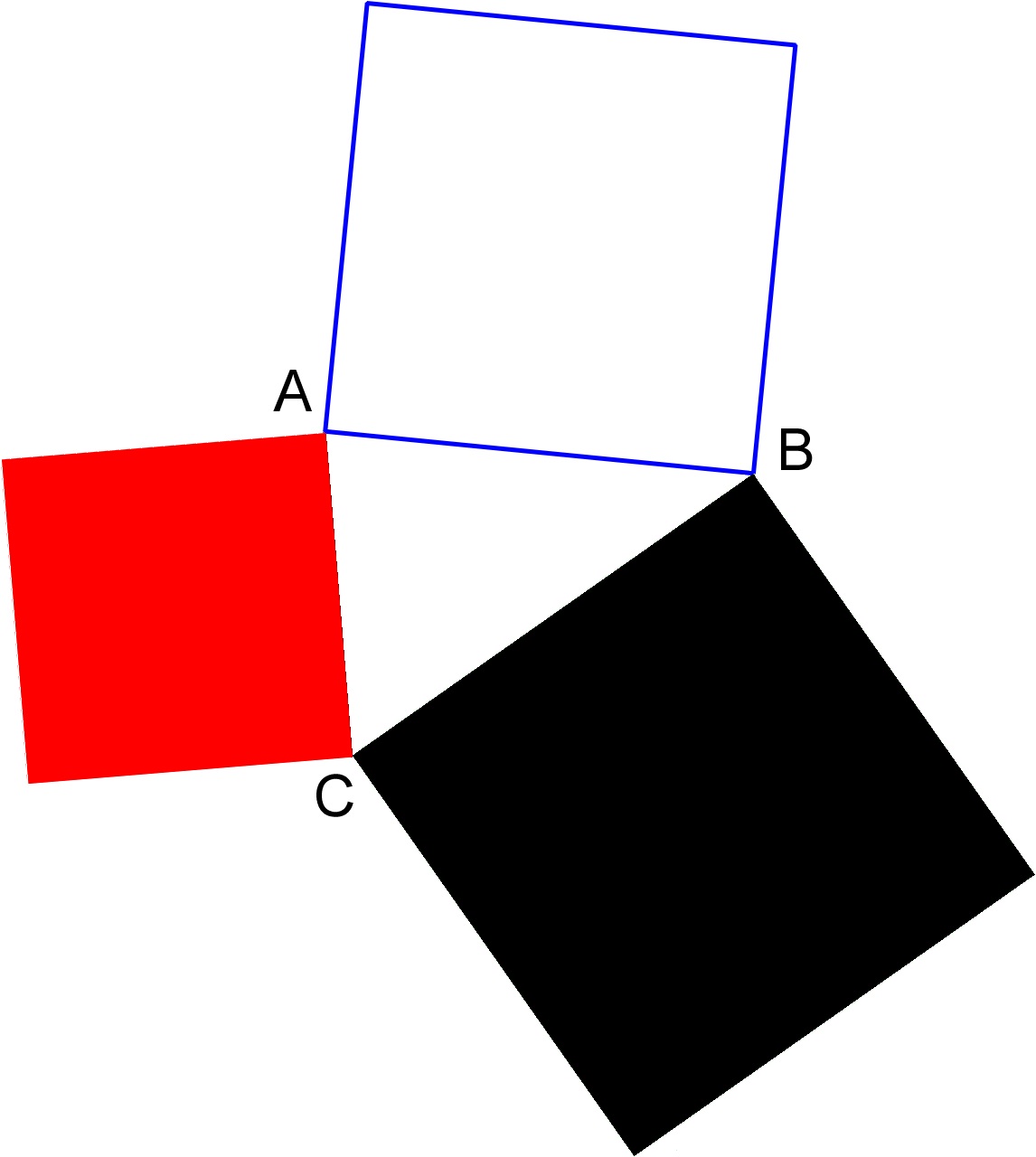}
 \includegraphics[width=50mm]{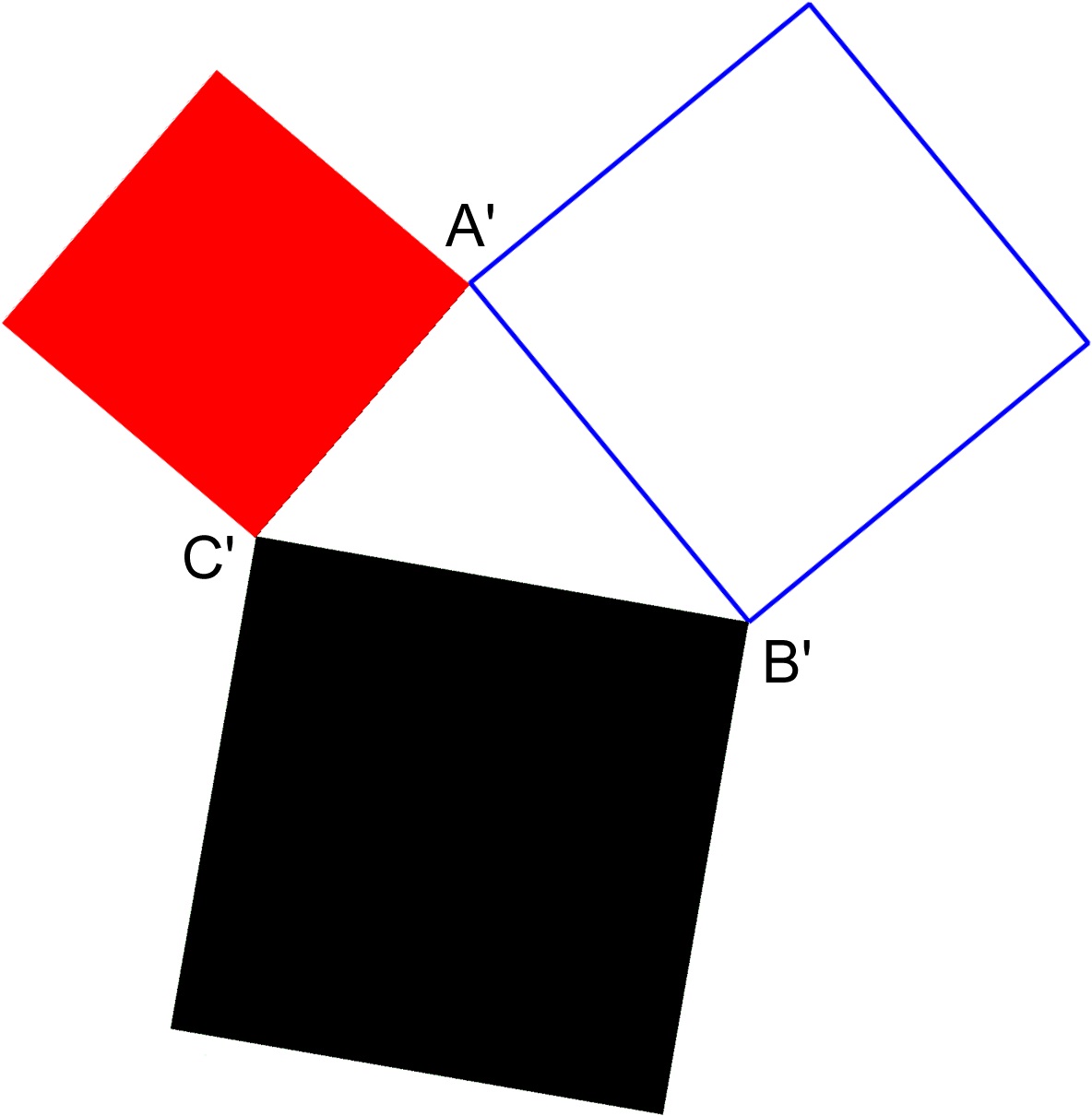} 
\includegraphics[width=50mm]{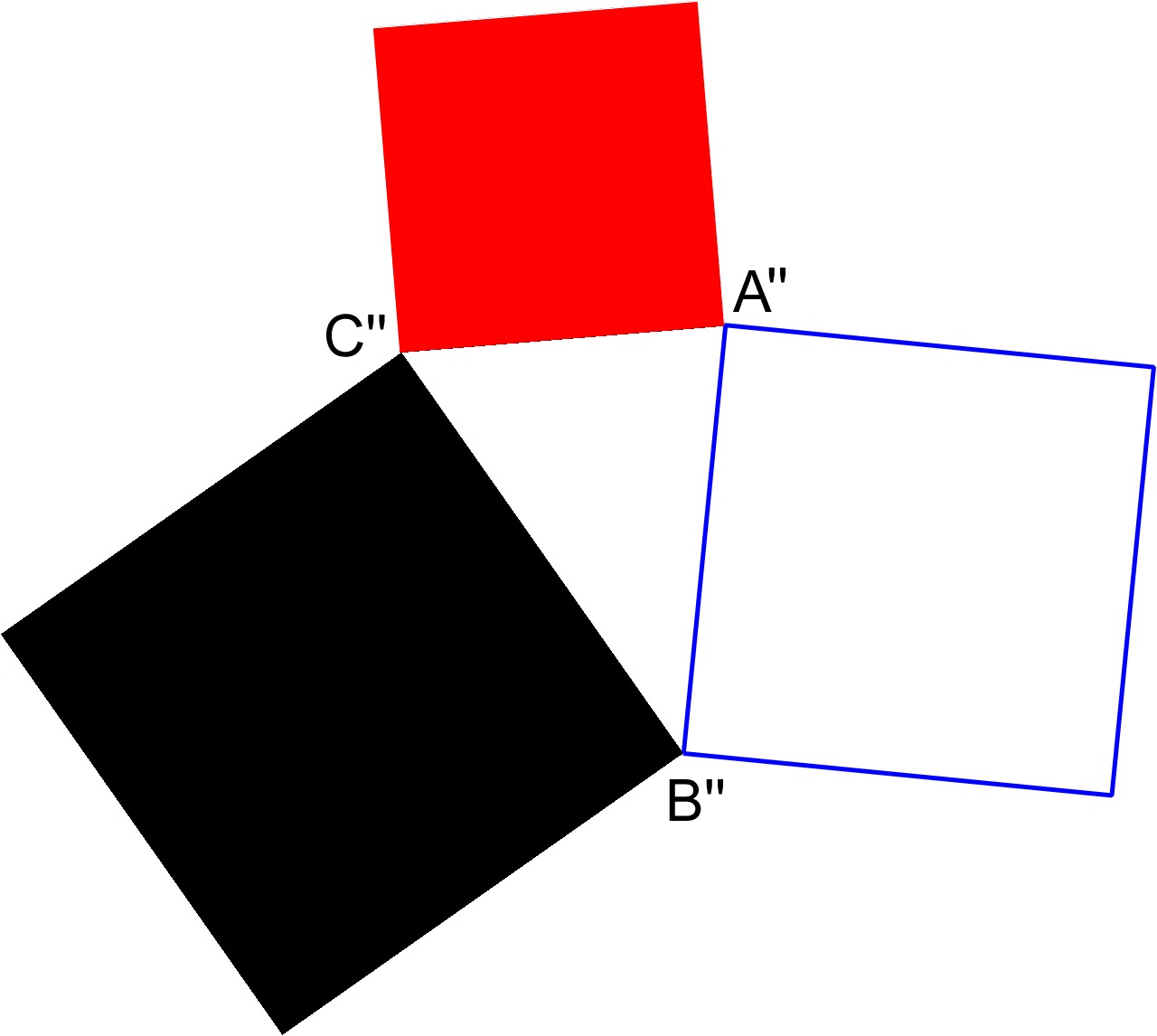}\\
\caption{ Three triadas of Malevich's squares describing the qutrit 
state and corresponding to three artificial qubit states, 
respectively.}
\end{figure}

\section{Quantum Channels for Qutrit States}
The linear maps of the qutrit-state density matrix (\ref{eq.17}) can
be expressed as a linear transform of the  eight-dimensional vector
$\vec \Pi$ with the components $p_1^{(1)}$, $p_2^{(1)}$,
$p_3^{(1)}$, $p_1^{(2)}$, $p_2^{(2)}$, $p_3^{(2)}$, $p_1^{(3)}$, and
$p_2^{(3)}$. These components are the probabilities for three
artificial spin-1/2 systems and three spin projections $m=1/2$ onto
three perpendicular directions in the space. In fact, we have also
the probabilities $p_3^{(3)}=p_3^{(2)}$. The unitary transform of
the density matrix $\rho$\\[-5mm]
\begin{equation}\label{eqA1}
\rho\longrightarrow\rho_u=u\rho u^{\dagger},
\end{equation}
where $u u^{\dagger}=1$, provides the linear transform of the
eight-vector $\vec\Pi$. One can get this transform (quantum channel)
in an explicit form.

In fact, the nine-dimensional vector $\vec\rho$ with components
$\left(\rho_{11},\rho_{12},\rho_{13},\rho_{21},\rho_{22},\rho_{23},
\rho_{31},\rho_{32},\rho_{33}\right)$,
which can be denoted as
$\left(\rho_{1},\rho_{2},\rho_{3},\rho_{4},\rho_{5},\rho_{6},\rho_{7},\rho_{8},\rho_{9}\right)$,
after the unitary transform converts to the vector
$\vec\rho_u=u\otimes u^\ast\vec\rho$. Then we arrive at\\[-8mm]
\begin{equation}\label{eqA2}
\Pi_k'=\sum_{j=1}^8\bar{U}_{k j}\Pi_j+\Gamma_k,
\end{equation}
where the 9$\times$9 unitary matrix $U=u\otimes u^\ast$ determines
the 8$\times$8 matrix $\bar{U}_{k j}$ and the eight-vector
$\vec\Gamma$.

Since the components of vectors $\vec\Pi$ and $\vec{\Pi'}$ are
expressed in terms of the probabilities $p_j^{(k)}$ and
$p_j^{(k)'}$, the channel under discussion provides an explicit
transform of the probabilities determining the qutrit density
matrices. For a unital channel of the form
$\rho\longrightarrow\rho_U=\sum_{k} p_k u_k\rho u_k^\dagger$ ($0\geq
p_k\geq0$, $\sum_{k}p_k=1$), the transform of the vector
$\vec\rho\longrightarrow\vec\rho_U$ reads\\[-6mm]
\begin{equation}\label{eqA3}
\vec\rho_U=\left(\sum_k p_k u_k\otimes u_k^\ast\right)\vec\rho.
\end{equation}
It provides the transform of the eight-vector
$\vec\Pi\longrightarrow\vec\Pi_U$ of the form \\[-5mm]
\begin{equation}\label{eqA4}
\vec\Pi_U=\left(\sum_k p_k \bar U_k\right)\vec\Pi+\vec\Gamma,
\end{equation}
where the eight-vectors $\vec\Pi$ and $\vec{\Pi}_U$ are expressed in
terms of eight probabilities determining the qutrit-state density
matrix $\rho$. An analogous relation describes the generic
completely positive map of the qutrit-state density matrix
\begin{equation} \label{eqA5}
\rho\longrightarrow\rho_{\rm pos}=\sum_k V_k\rho V_k^\dagger,
\end{equation}
where $V_k$ are arbitrary 3$\times$3 matrices satisfying the
relation 
$~\sum_k V_k^\dagger V_k=1$\,.

For example, the channel providing the transform $\rho_{k
k}\longrightarrow\rho_{k k}'=\rho_{k k}$ and $\rho_{k
j}\longrightarrow\rho_{kj}'=0$ ($k\neq j$) for the qutrit-state
density matrix determines the transform of the probabilities
$~p_1^{(k)'}=1/2$, $~p_2^{(k)'}=1/2$, $~p_3^{(k)'}=p_3^{(k)}$,
$k=1,2,3$. This means that such a channel transforms the states of
three artificial qubits determining the initial density matrix of
the qutrit state into the state with maximum entropy $S=\ln 2$ for
the probability distributions of spin-projections $m=\pm1/2$ along
the axes $x$ and $y$. The positive map can be determined by the
combination of the described maps with the transforms
$\rho(1)\longrightarrow\rho^{\rm tr}(1)$ and
$\rho(2)\longrightarrow\rho^{\rm tr}(2)$ of the two artificial
qubit-state density matrices, as well as an analogous transposition
of the third artificial qubit-state density matrix.

\section{Conclusions}   
To conclude, we point out the main results of this work.

We presented the matrix elements of the qutrit-state density matrix
as linear combinations of nine classical probabilities $p_1^{(1)}$,
$p_2^{(1)}$, $p_3^{(1)}$, $p_1^{(2)}$,  $p_2^{(2)}$, $p_3^{(2)}$,
$p_1^{(3)}$, $p_2^{(3)}=p_3^{(3)}$. We interpreted the probabilities
$p_j^{(k)}$, $j,k=1,2,3$ as the probabilities to have ``spin-1/2
projections'' $m=1/2$ in three perpendicular directions of three
artificial qubits. This means that such quantum system as qutrit has
states whose density matrices are given in the classical formulation
by eight independent parameters -- eight probabilities corresponding
to the states of eight classical coins.

We found new inequalities for the introduced classical
probabilities, including new entropic inequalities for the
qutrit-state density matrix elements. The new inequalities provide
the condition of quantumness of qutrit. The states of eight
classical coins are described by the same probabilities, but these
probabilities should not satisfy these constrains. The new relations
for the qutrit-state density matrices obtained can be checked in
experiments with superconducting circuits~\cite{circuits} based on
Josephson junctions, which have been discussed in connection with
the nonstationary (dynamical) Casimir effect
in~\cite{Josephson,JSLR1991,Korea,measurementtech}; see also recent
publications~\cite{Kiktenko,AIP,PhysScr,Kiktenko1,Martinis,Astafiev,Ustinov}.
The dynamical Casimir effect was discovered in~\cite{Nori} and
discussed in~\cite{Dodonov1,Dodonov2,Dodonov3,Dodonov4}.

We presented the observables of the spin-1/2 system in the form of
three classical random variables $\vec X$, $\vec Y$, and $\vec Z$,
which are described by classical probability vectors $\vec{\cal
P}_1$, $\vec{\cal P}_2$, and $\vec{\cal P}_3$. These three classical
variables are organized in the form of the Hermitian matrix. The
quantumness of the construction is reflected by the introduced
inequalities for the probability distributions. We extended an
analogous construction for qutrit states and observables. We
discussed quantum channels for qutrit states in the probability
representation. Some aspects of the quantum channel properties in
the tomographic-probability picture are presented in
\cite{Mendes,Amosov}. We considered the triangle geometry of qutrit
states and described the states by the three triadas of Malevich's
squares in the quantum suprematism approach (suprematism in the art
is reviewed in~\cite{kniga}). The consideration can be extended to
arbitrary systems of qudit states. We will study this problem in a
future publication.

\section*{Acknowledgments}  
The formulation of the problem of quantum suprematism and the
results of Sec.~2 are due to V.~I.~Man'ko, who is supported by the
Russian Science Foundation under Project No.~16-11-00084; the work
was partially performed at the Moscow Institute of Physics and
Technology.

\end{document}